\documentclass[12pt]{article}
\usepackage{amsmath}
\usepackage{amsfonts}
\usepackage{amssymb}
\usepackage{amstext}
\usepackage{graphicx}
\usepackage{enumerate}
\usepackage{epsfig}%
\title{\centerline
\bf Non-minimally coupled dark fluid in Schwarzschild spacetime}
\bigskip
\author{Barnali Das$^\dagger$, Kaushik Bhattacharya$^\ddagger$
\thanks{email:\,\,$^\dagger$barnali@ncra.tifr.res.in, 
$^\ddagger$kaushikb@iitk.ac.in}
\\
\normalsize
$\dagger$National Centre for Radio Astrophysics,\\
\normalsize
Tata Institute of Fundamental Research, Pune\\
\normalsize
Pune University Campus, Ganeshkhind Pune 411007, India\\
\normalsize
$\ddagger$Department of Physics, Indian Institute of Technology, Kanpur\\
\normalsize
Kanpur 208016, India
}
\begin{document}
\maketitle
\begin{abstract}
If one assumes a particular form of non-minimal coupling, called the
conformal coupling, of a perfect fluid with gravity in the
fluid-gravity Lagrangian then one gets modified Einstein field
equation. In the modified Einstein equation, the effect of the
non-minimal coupling does not vanish if one works with spacetimes for
which the Ricci scalar vanishes. In the present work we use the
Schwarzschild metric in the modified Einstein equation, in presence of
non-minimal coupling with a fluid, and find out the energy-density and
pressure of the fluid. In the present case the perfect fluid is part
of the solution of the modified Einstein equation. We also solve the
modified Einstein equation, using the flat spacetime metric and show
that in presence of non-minimal coupling one can accommodate a perfect
fluid of uniform energy-density and pressure in the flat spacetime. In
both the cases the fluid pressure turns out to be negative. Except
these non-trivial solutions it must be noted that the vacuum solutions
also remain as trivial valid solutions of the modified Einstein
equation in presence of non-minimal coupling.
\end{abstract}
\section{Introduction}

General relativity (GR) is now accepted as the standard theory of
gravity whose application involves a wide variety of scales. From the
cosmic scale to the scale of neutron stars GR works
efficiently. Except the vacuum solutions all the other scenarios in GR
requires the presence of a gravitating fluid whose energy density and
pressure curves space and time. The dynamics of our universe is often
described by considering it as a perfect fluid. A perfect fluid is one
that has isotropic pressure, no viscosity and no heat
conduction. However, the validity of the perfect fluid description
obviously depends on the length scale used by the observer. We can
imagine the constituent bodies to be analogous to those in a fluid if
and only if the mean free path between them is much smaller than the
length scale used. If the mean free path of the constituent particles
of the system is small compared to the curvature scale, we can
consider the fluid to be a perfect fluid (in the simplest
approximation) in a locally flat spacetime and can apply minimal
coupling (MC) prescription to write down the energy momentum tensor of
the fluid in more longer length scales where curvature effects become
prominent.  The root of the procedure lies in the Einstein's
equivalence principle (EEP), which states that in small enough regions
of spacetime the laws of physics reduce to those of special
relativity. To get the equations in curved spacetime, we simply need
to change the flat Lorentzian metric by the actual metric for the
curved spacetime and also replace the partial derivatives by the
covariant derivatives. This includes the assumption that in the matter
Lagrangian, there is no coupling between the dynamical variables
describing the fluid and the curvature tensor \cite{Goenner:1984}. In
presence of non-minimal coupling (NMC), which may arise if the mean
free path of the matter distribution is comparable or larger than the
curvature scales, we will see that EEP does not hold as new effects
become apparent in the locally flat spacetime limit\footnote{One can
  see how NMC of gravity and Electromagnetic fields affect the
  equivalence principle in Ref.\cite{Prasanna:2003ix}}.

Recent observations of the Universe \cite{Frieman:2008sn,
  Frieman:2002wi}, using information from the CMBR and type I
supernovae data, have revealed that only $4 \%$ of the total energy
density is in the form of baryonic matter. Rest of the $96 \%$
consists of dark energy ($73 \%$) and dark matter ($23 \%$), none of
which has yet been experimentally detected. In an earlier work
\cite{Bettoni:2015wla} the authors have suggested that the interesting
behavior of the dark sector of the universe may have a connection with
fluids non-minimally coupled to gravity.  The authors of 
Ref.~\cite{Bettoni:2015wla} have argued that dark matter system, being
weakly interacting, may have very long mean free path ($\sim$ 1000
GPc); so at the length scale at which these kind of matter can be
regarded as a fluid, spacetime no longer remains flat and the effect
of curvature starts affecting the fluid variables directly. In that
case, the system can be described as a perfect fluid in a curved
spacetime which is non-minimally coupled to gravity. As soon as we
consider NMC of matter with gravity, the assumption that dynamical
variables do not couple directly with the curvature tensor breaks down
and the EEP is also violated\footnote{The idea of dark matter fluid
non-minimally coupled to gravity is also addressed in several other
works as in Refs.~ \cite{Bruneton:2008fk}, \cite{Bettoni:2011fs} and 
\cite{Bettoni:2012xv}}.

To include the effects of NMC of fluids with gravity one has to use
the action of perfect fluids in presence of gravity. The concept of
relativistic perfect fluids and their properties are well described in
Ref.~\cite{Andersson:2006nr} and \cite{Weinberg:1972kfs}, but in these
references the authors do not use the action principle.  The action
principle for relativistic fluids in general and for MC fluid in a
gravitational field is well studied \cite{Schutz:1977df, Brown:1992kc,
  Taub:1954zz, Chiueh:1994zz}. The authors of
Ref.~\cite{Bettoni:2015wla} expanded the terms in the fluid action by
introducing new NMC terms which they call the conformal coupling,
where the fluid variables couple with the Ricci scalar. There can be
another kind of a NMC where the fluid variables couple to the
contracted form of the Ricci tensor with the fluid 4-velocities,
called the disformal coupling. In the present work we will only focus
on the conformal coupling case. The generalization of the results for
the disformally coupled case can be done in a separate work. An
interesting outcome of the NMC of a perfect fluid to curvature is that
the gravitational field equation gets modified. In absence of NMC one
obtains the conventional Einstein equation for the metric in presence
of the perfect fluid whereas in presence of NMC the Einstein equation
gets modified. The NMC alters the Planck mass and the energy momentum
tensor of the fluid which now becomes dependent on the curvature
scale. In such a scenario one can analyze the modification of the
cosmological equations and get interesting outcomes. One naturally
expects that the cosmological solutions will be modified as the fluid
content now becomes non-minimally coupled to curvature \cite{Bettoni:2015wla}.

In the present work we have tried to address a different problem which
may arise in theories where a perfect fluid is non-minimally coupled
to gravity. We know that in Einstein gravity the Schwarzschild
solution is a vacuum, Ricci-flat solution which is asymptotically
flat.  Ricci flat solutions in general do not admit presence of matter
and are vacuum solutions of Einstein's equation in GR.  We pose the
question, can a Ricci-flat metric admit of matter in the presence of
NMC terms? In the present article we discuss about two simple
Ricci-flat cases, the Minkowski spacetime and that of the
Schwarzschild spacetime.  The result we get is interesting as because
we see that except the standard Schwarzschild solution in vacuum the
altered Einstein equation, which includes the effect of the
non-minimal coupling, also allows presence of matter in Schwarzschild
spacetime. This result opens up another question regarding the
asymptotic flatness of the Schwarzschild spacetime. The Schwarzschild
spacetime is known to be asymptotically flat and consequently the
Schwarzschild solution in presence of a perfect fluid must also be
asymptotically flat. In GR the flat spacetime solution is devoid of
matter but can one obtain a solution of the altered Einstein equation
where a flat spacetime can also accommodate a perfect fluid?  If the
answer is yes then the Schwarzschild solution in presence of a perfect
fluid can match the asymptotically flat spacetime in presence of the
same perfect fluid. It turns out that when one includes the NMC term
in the basic gravity-fluid action one can obtain flat spacetime
solution with a fluid with uniform energy density. The conventional
vacuum solution remains a trivial alternative.  In presence of NMC of
fluid with curvature we will see that if one demands that the number
density of particles in the fluid and energy density of the fluid be
positive definite the pressure of the resultant fluid has to be
negative. Consequently we propose an interesting solution of the
altered Einstein equation in presence of NMC where the Schwarzschild
solution can accommodate a perfect fluid with negative pressure. It is
a natural question that how such a fluid can exist without a
backreaction which modifies the form of the metric? The answer lies in
the NMC term, in the fluid-gravity Lagrangian, which absorbs all the
back reaction which the fluid may produce. As because the pressure of
the perfect fluid in the Schwarzschild spacetime turns out to be
negative we assume it to be a dark fluid. In this scenario blackholes
may be immersed in a dark fluid which fills up all the space outside
the blackhole with a varying energy density and pressure.

Before ending the introduction we must point out that in GR one can
always accommodate a perfect fluid in the physically relevant spatial
regions in the Schwarzschild spacetime\cite{LeFloch:2015ziy,
  Guzman:2011jy, Guzman:2011ka, Bini:2012zzd} where the Einstein
equation $R_{\mu \nu}=0$ holds true. As Schwarzschild metric is a
vacuum solution in GR, consequently the perfect fluid in such a
spacetime must not be backreacting. Alternatively, the metric must
affect the fluid properties via the minimal prescription procedure but
the fluid cannot affect the Schwarzschild metric properties in GR. One
can apply the conservation of the energy-momentum tensor of the
perfect fluid in such a case and find out the energy density profile
of the test fluid \cite{Bini:2012zzd}. In the case of non-minimal
coupling one can always accommodate a perfect fluid in the
Schwarzschild spacetime where the energy-momentum tensor of the
perfect fluid naturally conserved as it is a part of the altered
Einstein equation. Once NMC effects are taken into account the fluid
present in the Schwarzschild metric need not be a test fluid.  In
section \ref{mincoup} we will present some properties of a minimally
coupled perfect fluid in the presence of Schwarzschild spacetime. The
result of that section will be used to make a comparison between the
properties of a non-minimally coupled fluid and a minimally coupled
fluid in the Schwarzschild spacetime.

The materials in this paper are presented in the following way. In the
next section we specify the formulation of the problem which we will
like to address in the present paper. The notations and conventions
will also be defined in the next section. The solution of the modified
Einstein equation in presence of NMC, for the simplest cases where
$R=0$, will be presented in section \ref{flatsp}. The first subsection
of section \ref{flatsp} will focus on the fluid solutions for the case
of flat spacetime. The second subsection of section \ref{flatsp} has
the main solution of the modified Einstein equation in Schwarzschild
spacetime. In both of the cases in section \ref{flatsp} we will find
out the properties of the fluids assuming the background metric to be
known. The case of a perfect fluid minimally coupled to gravity, in
the Schwarzschild spacetime, is presented in section
\ref{mincoup}. The penultimate section presents a brief overview on
the stability of fluid perturbations in the fixed Schwarzschild
metric. The last section concludes the article and contains the
summary of the main results obtained in this paper.
\section{Formulation of the problem} 
\label{form}

In this section we formulate the problem after describing the basic
equations and the conventions and notations we will use extensively in
this paper. We start with the definition of stress-energy tensor SET
(or the energy momentum tensor) denoted by $T_{\mu\nu}$ for a perfect
fluid in terms of its energy density $\rho$ and pressure $p$:
\begin{eqnarray}
T_{\mu\nu}=(\rho+p)u_{\mu}u_{\nu}+pg_{\mu\nu} \,.
\label{emt}
\nonumber
\end{eqnarray}
Here $u_{\mu}$ is the fluid 4-velocity which is a timelike 4-vector
normalized as $u^\mu u_\mu=-1$ and $g_{\mu\nu}$ is the metric
tensor. For a minimally coupled perfect fluid, the action can be
written as :
\begin{align}
S_{\rm fluid}&=\int
d^{4}x\sqrt{-g}F(n,s)+J^{\mu}(\nabla_{\mu}\phi+s\nabla_{\mu}
\theta+\beta_{A}\nabla_{\mu}\alpha^{A})\,.
\label{inact}
\end{align}
Here,$J^{\mu}=\sqrt{-g}nu^{\mu}$ is a vector density field, $n$ is the
proper number density and $s$ is the entropy per particle. $F(n,s)$ is
a scalar function later revealed to be the negative of the energy
density $\rho$. The other variables $\phi$ and $\theta$ are spacetime
scalars related to chemical free energy, local temperature of the
fluid\cite{Brown:1992kc}. Except these variables one also has the
Lagrangian coordinates of the fluid element $\alpha^{A}$ where $A$
runs from 1 to 3.  The $\beta_A$ are  space time scalars which
represents Lagrange multipliers responsible for constraining the fluid
4-velocity. In our notation $\nabla_\mu$ stands for the covariant
derivative and in our convention,
$$\nabla_\mu A^\nu = \partial_\mu A^\nu + \Gamma^\nu_{\mu \,\alpha}
A^\alpha\,,$$ where $A^\nu$ is an arbitrary 4-vector and
$$\Gamma^\nu_{\mu\,\alpha}=\frac12 g^{\nu \kappa}(\partial_\mu
g_{\alpha \kappa} + \partial_\alpha g_{\mu \kappa} - \partial_\kappa
g_{\mu \alpha})\,,$$ stands for the affine connection coefficient. Except the
first term $\sqrt{-g}F(n,s)$ all the other terms in the action
integral imposes some constraints on the fluid flow and do not appear
explicitly in the fluid SET. One can connect the terms in the action
with the SET by using the standard relation
$$T_{\mu\nu}=\frac{2}{\sqrt{-g}}\frac{\delta{S}}{\delta{g_{\mu\nu}}}\,.$$
If one associates the energy density and the pressure appearing in the
SET in the following way, 
\begin{eqnarray}
\rho=-F\,, \,\,\,{\rm and}\,\,\, p=-n\frac{\partial F}{\partial n}+F\,, 
\label{rhop}
\end{eqnarray}
then the action principle can correctly reproduce the terms in the SET
of a perfect fluid.

If one wants to accommodate non-minimal conformal coupling in the
gravity-fluid action then the resultant action can be written as:
\begin{align}
S&=\frac{{M_{\rm P}}^{2}}{2}\int d^{4}x
\sqrt{-g}\left[1+\alpha_{c}F_{c}(n,s)\right]R + S_{\rm fluid}
\label{nmcact}
\end{align}
Where $S_{\rm fluid}$ corresponds to the minimally coupled fluid
action as given in Eq.~(\ref{inact}) and $M_{\rm P}=1/\sqrt{8\pi G}$ gives the
reduced Planck mass, $G$ being the universal gravitational constant. Here
$F_{c}(n,s)$ is a function of the fluid variables $n$ and $s$. 

In the Eq.~(\ref{nmcact}) the Ricci scalar is represented by $R$.  By
varying the metric in the action given in Eq.~(\ref{nmcact}) one
obtains the modified Einstein equation:
\begin{align}
{M_*}^2G_{\mu\nu}&=T_{\mu\nu}^{\rm eff} \,,
\label{meqn}
\end{align}
where the effective Planck mass is given by
${M_*}^2={M_{\rm P}}^2(1+\alpha_cF_c)$ and the effective SET has the form
\begin{align}
T_{\mu\nu}^{\rm eff}&=g_{\mu\nu}F-h_{\mu\nu}n\left(\frac{\partial F}
{\partial n}+\alpha_{c}\frac{M_{\rm P}^2}{2}R\frac{\partial
  F_{c}}{\partial n}\right)
-\alpha_{c}M_{\rm P}^2 \left(g_{\mu\nu}g^{\alpha\beta}\nabla_{\alpha}\nabla_{\beta} 
F_{c}-\nabla_{\mu}\nabla_{\nu}F_{c}\right)\,. 
\label{SET_c}
\end{align}
Here $h_{\mu\nu}=g_{\mu\nu}+u_{\mu}u_{\nu}$ is the transverse
projection tensor.  In the present case the modified Euler equation
take the following form:
\begin{eqnarray}
n\Big(\frac{\partial{F}}{\partial{n}}&+&\alpha_{c}\frac{M_{\rm P}^{2}}{2}R
\frac{\partial{F_{c}}}{\partial{n}}\Big)\dot{u}^{\sigma}-h^{\sigma\alpha}
\nabla_{\alpha}\Big[F-n\frac{\partial{F}}{\partial{n}}-\alpha_{c}
\frac{M_{\rm P}^{2}}{2}\Big(n\frac{\partial{F_c}}{\partial{n}}-F_{c}\Big)
R\Big]\nonumber\\
&+&\alpha_{c}\frac{M_{\rm P}^{2}}{2}F_{c}h^{\sigma\alpha}\nabla_{\alpha}R=0\,.
\label{feq_c}
\end{eqnarray}
For a minimally coupled fluid, the Euler equation is given as:
\begin{align}
(\rho+p)u^{\mu}\nabla_{\mu}u^{\sigma}+{h^{\sigma\nu}}\nabla_{\nu}p=0\,. 
\label{feq_m}
\end{align}
If one defines
\begin{eqnarray}
\rho_{\rm tot} = -\left(F+\alpha_{c}\frac{M_{\rm P}^{2}}{2}R
F_{c}\right)\,,\,\,\,\,
p_{\rm tot} = n\frac{\partial{\rho_{\rm tot}}}{\partial{n}}-\rho_{\rm
  tot}\,, 
\label{rhopt}
\end{eqnarray}
then Eq.~(\ref{feq_c}) can also be written as
\begin{align}
(\rho_{\rm tot}+p_{\rm tot})u^{\mu}\nabla_{\mu}u^{\sigma} +
  h^{\sigma\nu}\nabla_{\nu} p_{\rm tot}
-\alpha_{c}\frac{{M_{\rm P}}^{2}}{2} F_c \,h^{\sigma\alpha}\,\nabla_{\alpha}R=0 \,, 
\label{feq_cm}
\end{align}
which serves as the modified Euler equation in presence of non-minimal
conformal fluid-gravity coupling \footnote{In
  Ref.~\cite{Bettoni:2015wla} the authors miss a negative sign in the
  expression of pressure in Eq.~(2.4) which has been corrected in the
  present work. We would like to point out that the definition of
  $\rho_{\rm tot}$ used in Eq.~(\ref{rhopt}) differs from its form in
  Ref.~\cite{Bettoni:2015wla} where the authors use $\rho_{\rm
    tot}=F+\frac{\alpha_c}{2}{M_{\rm P}}^2RF_c$ instead of the form
  given in Eq.~(\ref{rhopt}). The expression of the energy density
  used here is consistent as in the absence of NMC we do obtain
  $\rho_{\rm tot}=-F$ as one expects for a minimally coupled perfect
  fluid.  The sign convention adopted in the present paper is
  consistent with a metric signature \{-,+,+,+\}}.

Comparing Eq.~(\ref{feq_m}) and Eq.~(\ref{feq_cm}), we see that
NMC introduces an extra force term which depends on the gradient of
the curvature. Also, total energy density and total pressure becomes
directly dependent on curvature which is a signature of NMC.  In
\cite{Bettoni:2015wla}, the authors have shown that the effect of NMC
remains even in the weak field limit as the effective Poisson equation
(for the Newtonian potential, $\phi_N$) in such a limit becomes
\begin{align*}
\nabla^2\phi_{N}&=4\pi G(\rho-\alpha_c\nabla^2 F^\prime_c)\,,
\end{align*}
where $F^\prime_c=F_c/(4\pi G)$. This effect shows that EEP
is violated in presence of NMC.

The above description of the basic theory of NMC of a fluid with
spacetime curvature opens up new questions regarding the solution of
the effective Einstein equation as given in Eq.~(\ref{meqn}). In
Eq.~(\ref{meqn}), the left hand side contains the Einstein tensor
$G_{\mu \nu}$ and the right hand side contains the effective SET. One
can assume some appropriate forms of $F$ and $F_c$ and solve the effective
Einstein equation for the metric. On the other hand one may also use
the equation the opposite way, where one feeds in a known spacetime
metric to see what are the resultant forms of $F$ and $F_c$ which
arises in such a spacetime. Out of various spacetimes one important
class of spacetimes is the one for which the tensor $G_{\mu \nu}=0$,
which constitutes the class of vacuum solutions in GR. Does a vacuum
solution in GR remain a vacuum solution of the effective Einstein
equation in presence of NMC? To answer the question one must choose
some $g_{\mu \nu}$ for which $G_{\mu \nu}=0$ and Eq.~(\ref{meqn})
becomes $T_{\mu\nu}^{\rm eff}=0$, which can be written explicitly as:
\begin{eqnarray}
g_{\mu\nu}F-h_{\mu\nu}\,n \frac{\partial F}{\partial n}
-\alpha_{c}M_{\rm P}^2 \left(g_{\mu\nu}g^{\alpha\beta}\nabla_{\alpha}\nabla_{\beta} 
F_{c}-\nabla_{\mu}\nabla_{\nu}F_{c}\right)=0\,,
\label{tmunuz}
\end{eqnarray}
and solve it to see whether one gets some reasonable forms of $F$ and $F_c$.
In the above equation we have used the fact that $R=0$ for all those
solutions which have $G_{\mu \nu}=0$. In general when one works with
non-minimally coupled fluid one assumes $F$ and $F_c$ to be some known
functions of $n$ and $s$. In the present case we do not know the exact
$n$ and $s$ dependence of $F$ or $F_c$ and consequently one has to
find out how $F_c$ depends on $n$ and $s$. The discussions on the next
two sections show that in these cases one can uniquely find out $F$
once one knows the coordinate dependence of $F_c$. In general $F_c$
will be found out by solving a differential equation, an equation
which can be treated as a dynamical equation for $F_c$. As a
consequence in the present article the status of $F_c$ is more like a
dynamical scalar field whose solution gives us the form of $F$. To
treat $F_c$ as a dynamical scalar field one also has to take a
variation of $F_c$ in the action integral in Eq.~(\ref{nmcact}). The
resulting field equation one gets by varying $F_c$ comes out to be
$R=0$, the subsector in which we are working. So our way of treating
$F_c$ as an independent field works perfectly for all those spacetimes
where one has $R=0$. In the present article we will assume the fluid
to be non-thermal and assume $F$ and $F_c$ are independent of $s$.
The modified Einstein equation, as given in Eq.~(\ref{tmunuz}) will
yield $F_c$ and $F$ as functions of spacetime coordinates. Using the
forms of these functions one can always express $n$ as a function of
the spacetime coordinates. Once the spacetime dependence of $n$ is
known one can invert the relationship and express both $F$ and $F_c$
as functions of $n$. For the case of a non-minimally coupled fluid in
flat spacetime one gets $F_c$ to be a function of $n$ and position
coordinates, in such a way that $F_c$ becomes zero when $n$ vanishes. 

If the differential equation, Eq.~(\ref{tmunuz}), is satisfied for a
non-zero $F$ and non-zero $F_c$ then one can unambiguously say that
the vacuum solution in GR can accommodate a perfect fluid in presence
of NMC. In this article we will specifically deal with Schwarzschild
solution in GR, which is a known vacuum solution in GR, and try to see
whether a spacetime defined by the Schwarzschild solution can
accommodate a perfect fluid when one uses the Schwarzschild metric in
the modified Einstein equation. As the Schwarzschild solution is
asymptotically flat one has to also investigate whether the Minkowski
spacetime also admits some matter in the presence of non-minimal
coupling. In the next section we will start our analysis with the flat
Minkowski solution and then move on to the Schwarzschild solution.
\section{Effect of non-minimal coupling for the simplest cases
 where the Ricci scalar is zero}
\label{flatsp}

In this section we will describe how a NMC affects the modified
Einstein equation for those spacetimes which has $R=0$. We will start
with the simplest case, the flat spacetime, and then move on to the
next case of Schwarzschild spacetime. As the Schwarzschild spacetime
is asymptotically flat so the effect of NMC in flat spacetime becomes
important.
\subsection{Non-minimally coupled fluid in flat spacetime}

The line element for flat spacetime is given by:
\begin{eqnarray}
ds^2&=-dt^2+dx^2+dy^2+dz^2\,.
\label{lst}
\end{eqnarray}
As $R_{\mu\nu}=0$ and $R=0$, we have $G_{\mu\nu}=0$. Before
proceeding further, we will make minor changes to Eq.~(\ref{tmunuz}) in
terms of the symbols used. We will like to rewrite Eq.~(\ref{tmunuz})
as
\begin{eqnarray}
g_{\mu\nu}\tilde{F}-h_{\mu\nu}\,n \frac{\partial \tilde{F}}{\partial n}
-\alpha_{c}M_{\rm P}^2 \left(g_{\mu\nu}\Box \tilde{F}_{c}-
\nabla_{\mu}\nabla_{\nu}\tilde{F}_{c}\right)=0\,,
\label{tmunuz_m}
\end{eqnarray} 
where the previous symbols (corresponding to $F$ and $F_c$) are only
represented in different notations without any new physical
input. Here and henceforth
$\Box=g^{\alpha\beta}\nabla_{\alpha}\nabla_{\beta}$. This change in
notation will help us to write down the equations in a more compact
way. Now we introduce the variables $F$ and $F_c$ as
\begin{eqnarray}
F=\frac{\tilde{F}}{{M_{\rm P}}^2}\,,\,\,\,\,F_c=\alpha_c \tilde{F_c}\,. 
\label{ndef}
\end{eqnarray}
In terms of the new variables as defined above the modified Einstein
equation in flat spacetime becomes,
\begin{eqnarray}
g_{\mu\nu} F - (u_\mu u_\nu + g_{\mu\nu}) n
\frac{\partial F}{\partial n}
-(g_{\mu\nu} \partial_\alpha \partial^\alpha F_c -
\partial_{\mu} \partial_{\nu} F_c)=0\,,
\label{fst1} 
\end{eqnarray}
where the metric is as given in Eq.~(\ref{lst}) and the fluid
4-velocity is assumed to be $u_\mu=(-1,0,0,0)$. The perfect fluid is
assumed to be at rest in the flat four dimensional manifold. Assuming
$F$ and $F_c$ to be independent of time, the above equation yields
\begin{eqnarray}
F - \partial_i \partial_i F_c = 0\,,
\label{zero}
\end{eqnarray}
and 
\begin{eqnarray}
F-n\frac{\partial F}{\partial n}-\left(\partial_i \partial_i F_c-
\frac{\partial^2 F_c}{\partial x^2}\right)&=& 0\,, \\
F-n\frac{\partial F}{\partial n}-\left(\partial_i \partial_i F_c-
\frac{\partial^2 F_c}{\partial y^2}\right)&=& 0\,, \\
F-n\frac{\partial F}{\partial n}-\left(\partial_i \partial_i F_c-
\frac{\partial^2 F_c}{\partial z^2}\right)&=& 0\,, 
\end{eqnarray}
which are the $0-0$ and $i-i$ components of Eq.~(\ref{fst1}). Here $i$
runs from 1 to 3.  The last three equations imply
\begin{eqnarray}
\frac{\partial^2 F_c}{\partial x^2}=\frac{\partial^2 F_c}{\partial y^2}
=\frac{\partial^2 F_c}{\partial z^2}\,,
\end{eqnarray}
which immediately gives
\begin{eqnarray}
F_c = C(x^2+y^2+z^2)=Cr^2 \,,
\label{fc1}
\end{eqnarray}
where $C$ is a constant. From Eq.~(\ref{zero}) one can get the
form of $F$ as 
\begin{eqnarray}
F=6C \,.
\label{f1}
\end{eqnarray}
One can now use these forms of $F_c$ and $F$ in the above equations
and find out 
\begin{eqnarray}
n\frac{\partial F}{\partial n}=2C\,,
\label{ndfdn}
\end{eqnarray}
which shows that $(\partial F/\partial n)$ is non-zero if $n$ does not
vanish identically. In order to satisfy the fact that $F$ is a
constant with $r$, but $(\partial F/\partial n)$ is non zero, one must
have
\begin{eqnarray}
F(n)=\alpha n^{\frac{1}{3}}\,, 
\label{fn1}
\end{eqnarray}
which yields,
\begin{eqnarray}
n=\left(\frac{F(n)}{\alpha}\right)^3=\left(\frac{6C}{\alpha}\right)^3\,,
\label{fstn}
\end{eqnarray}
if Eq.~(\ref{ndfdn}) is satisfied. In the above equations $\alpha$ is
another numerical constant. In flat spacetime one can indeed have a
fluid, which is non-minimally coupled to curvature, only if the
density of the fluid particles remains a constant throughout
spacetime. In the present case the existence of this energy-density
does not curve spacetime as because the effect of it is completely
cancelled by the existence of non-zero $F_c$. 

In the present case we see that
\begin{eqnarray}
F_c(n,r)=\frac{\alpha n^{\frac13} r^2}{6}\,,
\label{fc2}
\end{eqnarray}
which is zero if $n$ vanishes, but it also depends upon the radial
coordinate $r$. The above equation shows that $F_c$  increases with
distance, at infinite distance this coupling blows up but the number
density $n$ remains always a constant. Also from Eq.~(\ref{rhopt}) we
can write the energy density and pressure of the fluid to be,
\begin{eqnarray}
\rho_{\rm tot} = -6C\,,\,\,\,\, p_{\rm tot}= 4C\,, 
\label{fsrhop}
\end{eqnarray}
The above relations show that we have to take $C<0$ so that
$\rho$ remains positive definite. To make the number density $n$ to be
positive definite we also require to have $\alpha<0$. Consequently one
can write the expressions of the energy density and pressure of the
fluid also as
\begin{eqnarray}
\rho_{\rm tot}= -\alpha n^{\frac13}\,,\,\,\,\, p_{\rm tot} =\frac{2\alpha
  n^{\frac13}}{3}\,,\,\,\,\,\alpha<0\,,\,\,\,\,n>0\,, 
\label{fsrhop1}
\end{eqnarray}
predicting 
\begin{eqnarray}
p_{\rm tot}=-\frac23 \rho_{\rm tot} \,.
\label{fseos}
\end{eqnarray}
The fluid which satisfies the modified Einstein equation in flat
spacetime must have negative pressure.  These expressions for the flat
spacetime will be used to verify the asymptotic limit of Schwarzschild
solution in presence of NMC.

The modified Einstein equation, in flat spacetime, in presence of NMC
can certainly have the above non-trivial and interesting solution
which to our understanding was not presented before. On the other hand
one can get the standard vacuum solution, as one gets in GR, by
choosing the constant $C=0$ in the above analysis. In that case the
energy density, pressure of the fluid and the number density of the
fluid constituents vanish identically.
\subsection{Non-minimally coupled fluid in the Schwarzschild spacetime}
\label{nmcsb}
In this section we analyze the solution of the modified Einstein
equation in presence of NMC using the Schwarzschild metric. In GR we
know the Schwarzschild metric is a solution of the Einstein equation
in vacuum. We want to first verify whether the modified Einstein
equation allow the Schwarzschild spacetime to have matter. Next we
will try to figure out the energy density and pressure of the fluid in
the Schwarzschild spacetime.

We write the line element for Schwarzschild spacetime as
\begin{eqnarray}
ds^2=-(1-\frac{2m}{\tilde{r}}) dt^2+\frac{d\tilde{r}^2}
{1-\frac{2m}{\tilde{r}}}+\tilde{r}^2(d\theta^2 + \sin^{2}\theta
d\phi^2)\,.
\label{schm}
\end{eqnarray}
The usage of $\tilde{r}$ instead of $r$ is deliberate and the reason
for choosing such a convention will become clearer very soon. As like
the flat spacetime analysis, we will write the basic equation,
Eq.~(\ref{tmunuz}) as,
\begin{eqnarray}
g_{\mu\nu}\tilde{F}-h_{\mu\nu}\,n \frac{\partial \tilde{F}}{\partial
  n} -\alpha_{c}M_{\rm P}^2 \left(g_{\mu\nu}\Box \tilde{F}_{c}-
\nabla_{\mu}\nabla_{\nu}\tilde{F}_{c}\right)=0\,, 
\label{ntmunuz}
\end{eqnarray}
where the tildes over the quantities $F$ and $F_c$ are there for
purely technical reasons and shortly we will redefine the quantities
as $F$ and $F_c$ in appropriate ways. We will work in the comoving
frame and hence $u^i=0,\,\, i=1,2,3$ and $u^0u^0=-1/g_{00}$. Now we
can redefine various quantities as,
\begin{eqnarray}
r = \frac{\tilde{r}}{m}\,,\,\,\,\,F= \alpha_c \tilde{F}_c\,,\,\,\,\,
F=\frac{\tilde{F}m^2}{M_{\rm P}^2}\,,
\label{redf}
\end{eqnarray}
using which we will write the following equations in this section. In
this discussion we will assume that $n$, $s$, $F$ and $F_c$ are
time-independent quantities. Using the above notation the $0-0$
component of the modified Einstein equation becomes
\begin{eqnarray}
F-\Box F_c - \frac{1}{(1-\frac{2}{r})}\nabla_0\nabla_0 F_c = 0\,,
\label{eq_2}
\end{eqnarray}
where the notation is defined as
$$\nabla_t\nabla_t F_c = -\frac{1}{m^2
  r^2}\left(1-\frac2r\right)\frac{\partial F_c}{\partial
  r}=\frac{1}{m^2}\nabla_0\nabla_0 F_c\,.$$ In the above case we use
the coordinate $\tilde{r}$ to calculate $\nabla_t\nabla_t F_c$ and
then express the result in terms of $r$, as a result of which a factor
of $1/m^2$ comes out. The double covariant derivatives in terms of the
dimensionless $r$ modulo the $1/m^2$ factor is defined as
$\nabla_0\nabla_0 F_c$. Suitably manipulating the other components of
the modified Einstein equation, in the Schwarzschild spacetime, one
gets:
\begin{eqnarray}
F-n\frac{\partial F}{\partial n}-\Box F_{c}+\left(1-\frac{2}{r}\right)
\nabla_1\nabla_1 F_c&=&0\,, 
\label{eq_1}\\
r^2\left(1-\frac{2}{r}\right)\nabla_1 \nabla_1 F_c-\nabla_2\nabla_2 F_c&=&0\,, 
\label{eq_3}\\
\nabla_3 \nabla_3 F_c-\sin^2\theta\nabla_2\nabla_2 F_c&=&0\,, 
\label{eq_4}
\end{eqnarray}
where the double covariant
derivative $\nabla_{\tilde{r}}\nabla_{\tilde{r}} F_c$ using coordinate
$\tilde{r}$ is related to the quantity $\nabla_1\nabla_1 F_c$ via the
following relation,
$$\nabla_{\tilde{r}}\nabla_{\tilde{r}} F_c=\frac{1}{m^2}\left[
\frac{\partial^2 F_c}{\partial r^2}+\frac{1}{r^2(1-\frac2r)}
\frac{\partial F_c}{\partial r}\right]= \frac{1}{m^2}\nabla_1
\nabla_1 F_c\,.$$ The other double covariant derivatives appearing in
the above equations are given as,
$$\nabla_\theta\nabla_\theta F_c=\nabla_2\nabla_2 F_c\,,\,\,\,
\nabla_\phi\nabla_\phi F_c=\nabla_3\nabla_3 F_c\,,$$ where one
calculates the quantities $\nabla_\theta\nabla_\theta F_c$ and
$\nabla_\phi\nabla_\phi F_c$ using $\tilde{r}$ and re-express the
result in terms of $r$ in $\nabla_2\nabla_2 F_c$ and $\nabla_3\nabla_3
F_c$. In these cases one does not get factors of $1/m^2$ due to the
change of variables.

Solving the modified Einstein equations, in the Schwarzschild
spacetime, we will get the functional forms of $F$, $F_c$ and $n$. If
these quantities turns out to be non-vanishing then there can be a
perfect fluid in the Schwarzschild spacetime in such a way that the modified
Einstein equation is satisfied. In the present case one starts with 
Eq.~(\ref{eq_4}) which can be written as
\begin{eqnarray}
\frac{\partial^2 F_c}{\partial \phi^2} - \sin^2 \theta 
\frac{\partial^2 F_c}{\partial \theta^2} + \sin \theta \cos \theta 
\frac{\partial F_c}{\partial \theta}=0\,.
\end{eqnarray}
There can be various kind of solutions of the above equation. By using
the method of separation of variables one gets the following type of solutions:
\begin{eqnarray}
F_c(r,\theta,\phi) =
\left\{
	\begin{array}{ll}
		f(r)\,, \\
		f(r)\cos\theta\,, \\
		f(r)\sin\theta\cos\phi\,, \\
                f(r)\sin\theta\sin\phi\,, \\
		f(r)\sin\theta\cos\phi\log(\tan(\frac{\theta}{2}))\,,
	\end{array}
\right.
\end{eqnarray}
where the first solution is an obvious one. More over the first
solution can be uniquely chosen keeping spherical symmetry in mind.
Consequently we discard those solutions where $F_c$ explicitly depend
on $\theta$ and $\phi$. The only possible spherically symmetric
solution of $F_c$ must be a function of $r$ alone.
 
To find out the form of $f(r)$ we will use Eq.~(\ref{eq_3}) which can
be written explicitly as:
\begin{eqnarray}
r(r-2)\frac{d^2f}{dr^2} + (3-r)\frac{df}{dr}=0\,,
\end{eqnarray}
which can be transformed into a first order differential equation by
using a new variable $g(r)$ defined as $g(r)=df/dr$. The resulting first order
equation is
\begin{eqnarray}
\frac{dg}{g}+\left[-\frac{3}{2r}+\frac{1}{2(r-2)}\right]dr=0\,,
\end{eqnarray}
which after integration yields the result,
\begin{eqnarray}
\frac{dF_c}{dr}=A\sqrt{\frac{r^3}{|r-2|}}\,,
\end{eqnarray}
where $A$ is an integration constant. One can integrate the above
expression and finally obtain
\begin{eqnarray}
F_c(r) =
\left\{
	\begin{array}{ll}
		\tilde{B}+\tilde{A}\left[3\sin^{-1}\sqrt{\frac{r}{2}}-\frac{3}{2}
                \sqrt{r(2-r)}\right.\\
	\left.	-\frac{1}{2}\sqrt{r^3(2-r)}\right]\,;\,\,\, r<2\,, \\\\
		B+A\left[\frac{1}{2}\sqrt{r^3(r-2)}+\frac{3}{2}\sqrt{r(r-2)}
                 \right.\\
	\left. +3\log\left(\sqrt{\frac{r}{2}}+\sqrt{\frac{r-2}{2}}\right)
        \right]\,;\,\,\, r>2\,, 
	\end{array}
\right.
\end{eqnarray}
where the solutions for the regions $r>2$ and $r<2$ are separately specified.
Since $r<2$ is not physically accessible we will always take $r>2$ and
hence the physically relevant solution is
\begin{eqnarray}
F_c = B+A\left[\frac{1}{2}\sqrt{r^3(r-2)}+\frac{3}{2}\sqrt{r(r-2)}+ 
3\log\left(\sqrt{\frac{r}{2}}+\sqrt{\frac{r-2}{2}}\right)\right]\,,\,\,\,r>2\,.
\label{finfc}
\end{eqnarray}
From Eq.~(\ref{eq_2}) one can now find out the expression of $F$ as
\begin{eqnarray}
F&=&\frac{3(r-2)}{r^2}\frac{dF_c}{dr}\\
 &=& 3A\sqrt{\frac{r-2}{r}}\,,
\label{finf}
\end{eqnarray}
showing that $F$ tends to zero as $r$ approaches the value 2, near the
event horizon. From the last equation one can easily calculate
\begin{eqnarray} 
\frac{dF}{dr}=\frac{3}{r^3}\frac{dF_c}{dr}\,,
\label{aux1}
\end{eqnarray}
a result which will be useful for the calculation of $n$ as a function
of $r$. Using the above results in Eq.~(\ref{eq_1}) one gets
\begin{eqnarray}
n \frac{\partial F}{\partial n} = \frac{(r-3)}{r^2}\frac{dF_c}{dr}\,.
\label{aux2}
\end{eqnarray}
Writing the last equation as,
\begin{eqnarray}
n \frac{(dF/dr)}{(dn/dr)} = \frac{(r-3)}{r^2}\frac{dF_c}{dr}\,,
\end{eqnarray}
and using the result of Eq.~(\ref{aux1}) one gets the functional form
of $n$ as
\begin{eqnarray}
n(r)=D\frac{|r-3|}{r}\,, 
\label{nr}
\end{eqnarray}
where $D$ is a numerical constant. This shows that $n(r)$ is not a
constant in the finite $r$ region of the Schwarzschild spacetime. More
over $n(r)$ vanishes at $r=3$ and remains finite at $r=2$. 
\begin{figure}[t]
\begin{center}
\includegraphics[width=150mm]{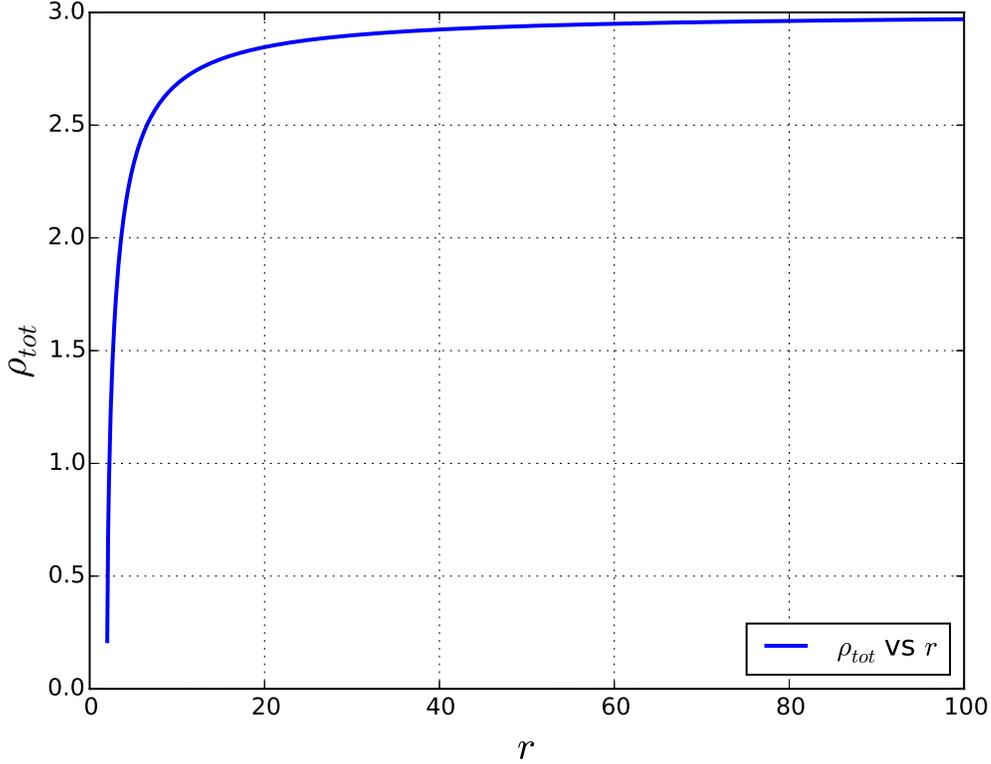}
\caption{Plot of $\rho_{\rm tot}$ with $r$.}
\label{fig:rhr}
\end{center}
\end{figure}

We can now calculate all the properties of the perfect fluid in the
Schwarzschild spacetime. In the present case if we want to make
the energy-density of the fluid to be positive then the constant $A$
has to be smaller than zero. Using Eq.~(\ref{rhopt}) we get
\begin{eqnarray}
\rho_{\rm tot} = -3A\sqrt{\frac{r-2}{r}}\,,\,\,\,\,
p_{\rm tot} = A \frac{(2r-3)}{r}\sqrt{\frac{r}{r-2}}\,,\,\,\,\,\,A<0\,,
\label{schrp}
\end{eqnarray}
which yields
\begin{eqnarray}
\frac{p_{\rm tot}}{\rho_{\rm tot}}=-\left(\frac{2r-3}{3r-6}\right)\,,
\label{scheos}
\end{eqnarray}
which reproduces the result in flat spacetime, given in
Eq.~(\ref{fseos}), in the limit of large $r$. In this case also we see
that for positive definite energy density one must have a negative
pressure for finite values of $r$. Consequently the perfect fluid
which is non-minimally coupled to gravity cannot have a positive
pressure and acts like an exotic dark fluid. From the expressions of
the fluid variables one sees that $\rho_{\rm tot} \to 0$ as $r
\to 2$ from the right hand side. On the other hand $p_{\rm tot} \to
-\infty$ as  $r\to 2$ from the right hand side. Consequently one finds
that
\begin{eqnarray}
-\infty < \left(\frac{p_{\rm tot}}{\rho_{\rm tot}}\right) < -\frac{2}{3}\,,
\label{limeos}
\end{eqnarray}
where the limiting values, of the equation of state of the fluid, are
attained at $r=2$ and $r \to \infty$. The equation of state of the
fluid is a function of $r$ and it increases as $r$ increases. At $r=3$
the equation of state becomes $(p_{\rm tot}/\rho_{\rm tot})=-1$ and in
between $r=2$ and $r=3$ the equation of state decreases from $-1$ to
$-\infty$. The phantom divide \cite{Caldwell:1999ew, Caldwell:2003vq}
happens at $r=3$.
\begin{figure}[t]
\begin{center}
\includegraphics[width=150mm]{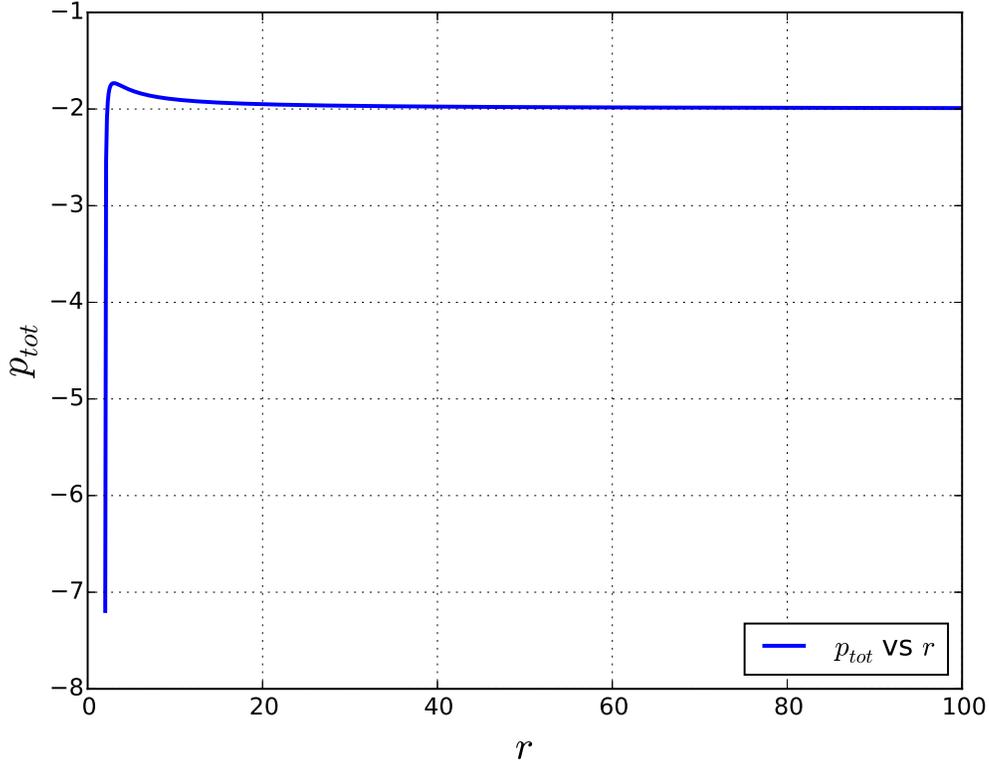}
\caption{Plot of $p_{\rm tot}$ with $r$.}
\label{fig:pr}
\end{center}
\end{figure}
The plots of $\rho_{\rm tot}$ and $p_{\rm tot}$ with respect to $r$
are presented in Fig.~\ref{fig:rhr} and Fig.~\ref{fig:pr}. The value of $A$ has
been taken as minus one. The specific choice of $A$ is justified in
Eq.~(\ref{setcns}). The value of $A$ is related to the large $r$
asymptotic values of $\rho_{\rm tot}$ and $p_{\rm tot}$.

One can express all the physical fluid parameters in terms of $n$. As
there is only one fluid in question and we do not take into account
thermal equilibrium, entropy density $s$ does not play a part in this
discussion. From Eq.~(\ref{nr}) one sees that
\begin{eqnarray}
r=\frac{3D}{D-n}\,,\,\,\,\,r>3\,,
\label{rgr3}
\end{eqnarray}
which directly shows that the constant $D$ should be such that $D>n$
for $r>3$. One can now express the energy-density in terms of $n$ as
\begin{eqnarray}
\rho_{\rm tot}(n) = -3A \left[\left(\frac{2}{3D}\right)n +
    \frac13\right]^{1/2}\,,\,\,\,\,\,r>3\,.
\label{rhogr3}
\end{eqnarray}
Working similarly one can get,
\begin{eqnarray}
\rho_{\rm tot}(n) = -3A \left[\frac13-
\left(\frac{2}{3D}\right)n\right]^{1/2}\,,\,\,\,\,\,2<r<3\,,
\label{rhol3}
\end{eqnarray}
which again restrains $D$ in this region as here one requires $D>2n$
for a real value of $\rho_{\rm tot}(n)$. The above results show that
although the radial derivative of $n(r)$ is a discontinuous function of
$r$ at $r=3$, where $n=0$, $\rho_{\rm tot}(n)$ remains a continuous
function of $n$ in the whole region of physical relevance.  One can
also express pressure as a function of $n$ in the various regions as:
\begin{eqnarray}
p_{\rm tot}(n)=
A\left(1+\frac{n}{D}\right)\left[\frac{3}{1+(\frac{2}{D})n}
\right]^{1/2}\,,\,\,\,\,r>3\,,
\label{pgr3}
\end{eqnarray}
and 
\begin{eqnarray}
p_{\rm tot}(n)=
A\left(1-\frac{n}{D}\right)\left[\frac{3}{1-(\frac{2}{D})n}
\right]^{1/2}\,,\,\,\,\,2<r<3\,,
\label{pl3}
\end{eqnarray}
which again shows that pressure varies continuously with $n$. From the
above expressions one can immediately notice that $\rho_{\rm tot} +
p_{\rm tot}>0$ for $r>3$ and $\rho_{\rm tot} + p_{\rm tot}<0$ for $2<r<3$.
\begin{figure}[ht]
\begin{center}
\includegraphics[width=150mm]{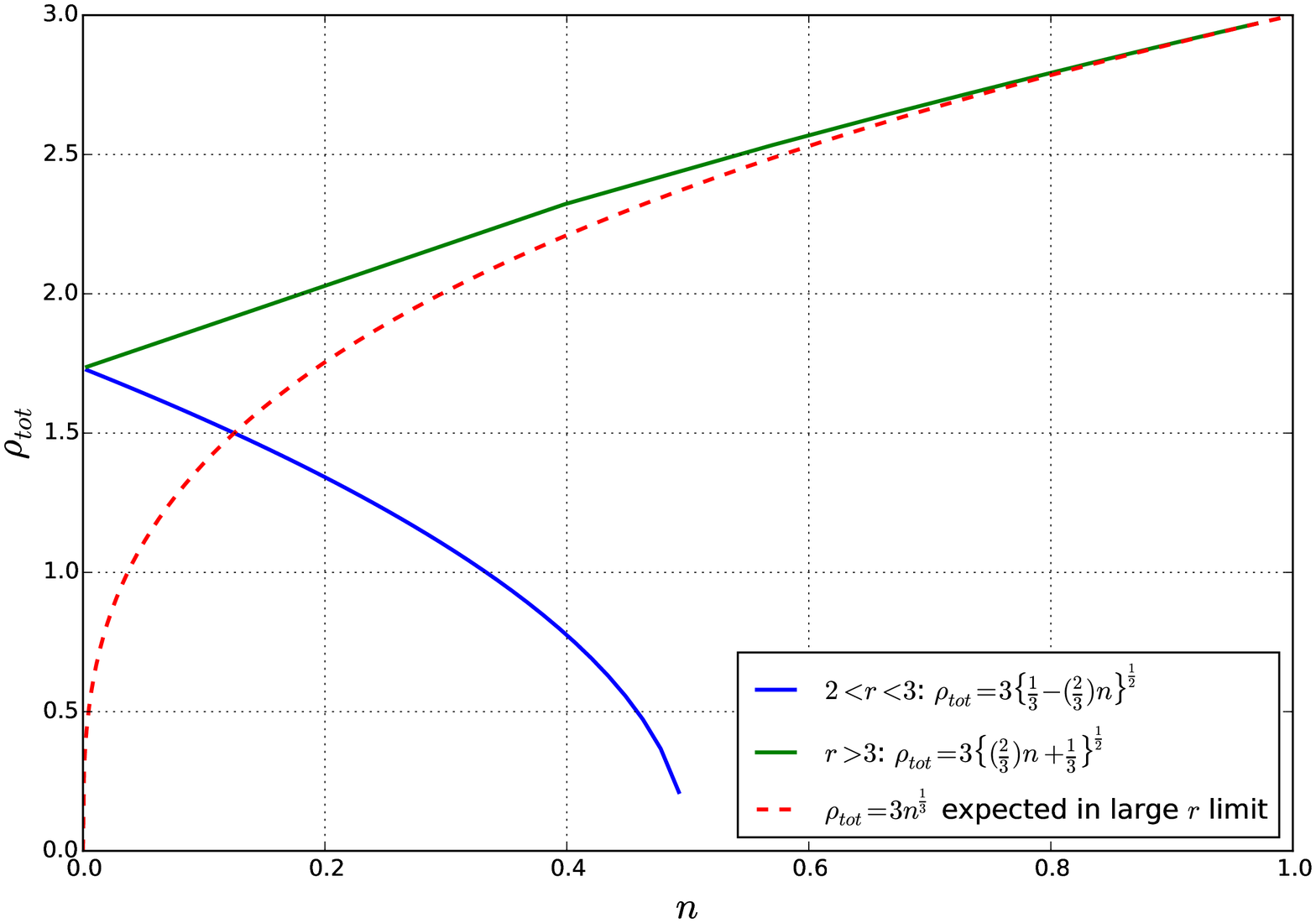}
\caption{Plot of $\rho_{\rm tot}$ with $n$.}
\label{fig:rn}
\end{center}
\end{figure}
From the form of Eq.~(\ref{ndfdn}) and Eq.~(\ref{nr}) one can identify
that
\begin{eqnarray}
D=\left(\frac{6C}{\alpha}\right)^3\,,
\label{match1}
\end{eqnarray}
as the Schwarzschild result tends to the flat spacetime result in the
large $r$ limit. On the other hand if one matches the value of the
energy density in Schwarzschild spacetime for large $r$ values to that
of the flat spacetime result in Eq.~(\ref{fsrhop}) one gets
$C=A/2$. Using this result in the last equation one gets a
relationship between the various constants as
\begin{eqnarray}
D=3A\left(\frac{36C^2}{\alpha^3}\right)\,.
\label{match2}
\end{eqnarray}
\begin{figure}[ht]
\begin{center}
\includegraphics[width=130mm]{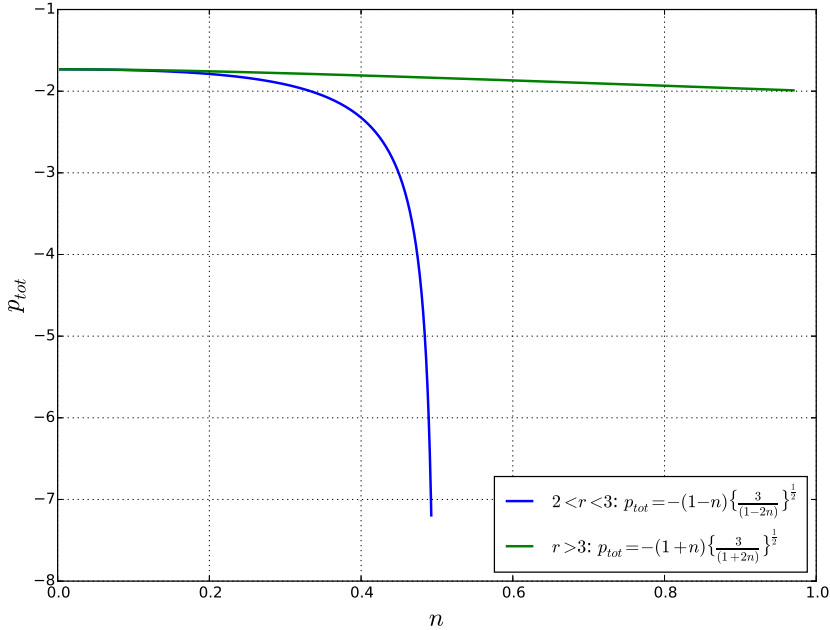}
\caption{Plot of $p_{\rm tot}$ with $n$.}
\label{fig:pn}
\end{center}
\end{figure}
The above equation connects the Schwarzschild spacetime results with
the flat spacetime results. As such the above equation does not
specify any particular set of values of the constants and may be valid for
a range  of values for the constants, giving various fluid profiles. In
this article we will choose a specific set of values of the constants
as
\begin{eqnarray} 
A=-1\,,\,\,\,\,D=1\,,\,\,\,\,C=-\frac12\,,\,\,\,\,\alpha=-3\,,
\label{setcns}
\end{eqnarray}
for which Eq.~(\ref{match2}) holds. One can always assume $B=0$, in
the expression of $F_c$ in Eq.~(\ref{finfc}), without any loss of
generality.  For finite values of $r$ one can always express $F_c(n)$,
in Eq.~(\ref{finfc}), as a function of $n$ using the functional
relationship between $n$ and $r$. For very large values of $r$
something interesting happens. For large enough $r$, one gets $F_c
\sim \frac{A}{2}r^2$ from Eq.~(\ref{finfc}). In this large $r$ limit
one cannot simply express $F_c$ as purely a function of $n$ as $n$
becomes constant for large $r$.  Remembering that $C=A/2$, one sees
that in the large $r$ limit one gets $F_c \sim C r^2$ which is exactly
similar to the relation in Eq.~(\ref{fc1}), a result familiar from our
flat spacetime analysis.

The plots of $\rho_{\rm tot}$ and $p_{\rm tot}$ with respect to $n$
are presented in Fig.~\ref{fig:rn} and Fig.~\ref{fig:pn}. Looking at
the plots one may like to think that both $\rho_{\rm tot}$ and $p_{\rm
  tot}$ are double valued functions of $n$ for some values of $n$, but
that is not the case here. In the figures we have specified the
various radial regions for the different branches. The number density
vanishes at $r=3$, where both the branches meet continuously. The
parameters required to draw the plots are taken from Eq.~(\ref{setcns}). 
\section{Minimally coupled fluid in the Schwarzschild spacetime}
\label{mincoup}

To understand the particular properties of non-minimally coupled fluid
in Schwarzschild spacetime it may be helpful to contrast the results
with those of the minimally coupled fluid in the Schwarzschild
spacetime. In this case the fluid is not backreacting on the spacetime
and consequently the Schwarzschild spacetime remains unaltered in the
presence of the fluid.  For the case of minimal coupling one has the
SET of the fluid as given in Eq.~(\ref{emt}) in section \ref{form}. To
maintain energy-momentum conservation one has to assume
\begin{eqnarray}
\nabla_\mu T^{\mu \nu}=0\,.
\label{emc}
\end{eqnarray}
To compare the two cases, in this case also we assume that $u_{\mu}$
is the fluid 4-velocity which is a timelike 4-vector is normalized as
$u^\mu u_\mu=-1$ and $g_{\mu\nu}$ is the Schwarzschild spacetime
metric as given in Eq.(\ref{schm}).  Assuming time-independent $\rho$
and $p$ we see that $\nabla_\mu T^{\mu 0}=0$ is an identity and
reveals no information regarding the fluid variables. The two
equations, $\nabla_\mu T^{\mu \theta}=0$ and $\nabla_\mu T^{\mu
  \phi}=0$ yield the important relations
\begin{eqnarray}
\left(\frac{\partial p}{\partial \theta}\right)=0\,,\,\,\,\,
\left(\frac{\partial p}{\partial \phi}\right)=0\,,
\label{tpind}
\end{eqnarray}
which says that for the perfect fluid at rest in presence of the
Schwarzschild spacetime, the pressure must be purely a function of
$r$. From the equation $\nabla_\mu T^{\mu \tilde{r}}=0$ one gets
\begin{eqnarray}
\frac{dp}{dr}+\frac{\rho+p}{r^2(1-\frac2r)}=0\,,
\label{rpsch}
\end{eqnarray}
where $r=\tilde{r}/m$. The above result shows that one cannot
accommodate a pressure-less, static perfect fluid, minimally coupled to
gravity, in the Schwarzschild spacetime. If one assumes that the
equation of state of the perfect fluid, minimally coupled to gravity,
is simply given by $p=\omega \rho$ then the solution of the above
differential equation is
\begin{eqnarray}
\rho(r)=\rho_0 \left(1-\frac2r\right)^{-\frac{(1+\omega)}{2\omega}}\,,
\label{eospfm}
\end{eqnarray}
where $\rho_0$ is a constant energy-density at very large $r$. In this
simple case one sees that if $\omega=0$ then $\rho(r)=0$ for $r>2$,
a fact we already know. On the other hand if $\omega=-1$ then
$\rho(r)$ is constant throughout the region $r>2$, which is the
familiar property of vacuum energy. If $\omega=1/3$ then the energy-
density (as well as pressure) is maximum near the event horizon and
the energy-density decreases as one moves away to large values of
$r$. If $\omega \ll -1$ then the energy-density is again
maximum near the event horizon but the pressure is negative and dips
to negative infinity near the event horizon. Only if $-1<\omega<0$
then $\rho(r)$ tends to zero as $r$ nears the event horizon and
pressure remains negative always. 

From the above discussion we see that the difference between
non-minimally coupled fluid and minimally coupled fluid in
Schwarzschild spacetime is remarkable. The most important difference
is related to the status of the equations which describe the nature of
these fluids. In the case of minimally coupled fluid, the SET was
assumed to be present and one had to invoke the energy-momentum
conservation equation for the fluid to mathematically formulate the
properties of such a fluid. For the case of NMC fluid, the equations
governing the nature of the fluid directly came from the modified
Einstein equation. The SET automatically satisfied energy-momentum
conservation condition as it was a part of the modified Einstein
equation. The non-minimally coupled fluid has a specific
energy-density and pressure in the Schwarzschild spacetime where as
the minimally coupled fluid only gives a relationship between $\rho$
and $p$ and nothing more. If one does not assume a barotropic fluid
then one has to find out the pressure of the fluid by assuming various
forms of $\rho(r)$. All the analysis in this paper is based on the
assumption of a static fluid and we hope there will be more
interesting properties of these fluids if they have non-zero
3-velocity components. In the next section we will briefly opine on
the effect of fluid velocity perturbations on the fluid properties 
when the fluid is non-minimally coupled to gravity.
\section{A brief note on the effect of perturbations of the fluid
  velocity}
\label{perturbs}
One can study the stability of the solution so obtained by
perturbative methods. There can be two kinds of perturbations involved
in the present case. In the first kind, we perturb the fluid
4-velocity keeping the background metric fixed. In the second kind,
one can perturb the metric itself and study the properties of
non-minimally coupled fluids in perturbed Schwarzschild spacetime. The
second case leads to a new problem where one has to solve for the
fluid properties in an altered metric. In this section we will opine
on perturbations of the first kind where the fluid velocity is
perturbed and set up the perturbation equations up to the first order
keeping the background metric fixed.

In subsection \ref{nmcsb} we have the solutions of the field equation,
Eq.~(\ref{ntmunuz}) in the static case, where
$$u^0 \ne 0\,,\,u^i=0\,.$$ 
In the case of the static fluid in the Schwarzschild spacetime we had
$u^0 u^0 = -1/g_{00}$.  Now let there be a finite ,
but small perturbation in the velocity of the fluid element
\begin{eqnarray}
U^0 = u^0(r) + \delta u^0(t,{\bf x})\,,\,\,\,U^i=
\delta u^i(t,{\bf x})\,, 
\label{inpert}
\end{eqnarray}
where $U^\mu(t,{\bf x})$ are the new perturbed 4-velocity
components. From the normalization of $U^\mu$ one can check that
$\delta u^0(t,{\bf x})=0$ up to first order in the perturbation. Due
to the perturbation in the fluid velocity all of the terms in
Eq.~(\ref{ntmunuz}) now becomes time-dependent. One must first note that the
fluid velocity enters the dynamical equations directly via $h_{\mu
  \nu}$. The diagonal terms in $h_{\mu \nu}$ does not change up to
first order of perturbation in the fluid 4-velocity, and consequently
the equations obtained from the diagonal terms in Eq.~(\ref{ntmunuz}) must
remain the same. Consequently the perturbations on $F(t,{\bf x})$,
$F_c(t,{\bf x})$ and $n(t,{\bf x})$ must also satisfy the old equations:
\begin{eqnarray}
\delta F-\Box \delta F_c - \frac{1}{(1-\frac{2}{r})}\nabla_0\nabla_0 \delta
F_c &=& 0\,,
\label{peq_1}\\
\delta F -\delta \left(n\frac{\partial F}{\partial n}\right)-
\Box \delta F_{c}+\left(1-\frac{2}{r}\right)
\nabla_1\nabla_1 \delta F_c &=& 0\,, 
\label{peq_2}\\
r^2\left(1-\frac{2}{r}\right)\nabla_1 \nabla_1 \delta F_c-
\nabla_2\nabla_2 \delta F_c &=& 0 \,, 
\label{peq_3}\\
\nabla_3 \nabla_3 \delta F_c-\sin^2\theta\nabla_2\nabla_2 \delta F_c &=& 0\,. 
\label{peq_4}
\end{eqnarray}
In our notation
$$F(t,{\bf x})=F_0(r) + \delta F(t,{\bf x})\,,\,\,
F_c(t,{\bf x})=F_{c0}(r) + \delta F_c(t,{\bf x})\,,\,\,
n\frac{\partial F}{\partial n}=\left(n \frac{\partial F}{\partial
  n}\right)_0 + \delta \left(n\frac{\partial F}{\partial n}\right)\,,$$
where the subscript zero stands for the unperturbed values of the
quantities.

The above equations do not show any dependence of the perturbed
quantities on the 3-velocity perturbations. To get the dependence of
the perturbed energy density, non-minimal coupling and the number
density one has to write the non-diagonal terms in
Eq.~(\ref{ntmunuz}). Up to first order in the perturbation of the fluid
velocity one must note that
$$\delta h_{ij}=0\,,\,\,\delta h_{0i}=u_0 \delta u_i\,,$$
where $i,j$ runs from one to three. Consequently the non-diagonal
terms of Eq.~(\ref{ntmunuz}) gives
\begin{eqnarray}
u_0 \delta u_i \left(n \frac{\partial F}{\partial
  n}\right)_0 -  \nabla_0 \nabla_i \delta F_c =0\,.
\label{nondiag}
\end{eqnarray}
The interesting thing to observe is that up to first order of
perturbations in the fluid velocity the diagonal equations keeps the
same form as those of the unperturbed solutions, although the
perturbations are now functions of space and time. The non-diagonal
equations specify the velocity perturbations in terms of the
non-minimal coupling. This observation is only true to the first order
of perturbations in the fluid velocity, in the second order of
perturbations the diagonal equations will be directly affected by the
velocity perturbations.

The way to solve the above set of equations for the perturbations is a
complicated matter as these equations can have non-trivial time
dependence. In this article we will not discuss the details about the
general method of solving such equations, it is beyond the scope of
the present paper. We will end this brief note on the fluid
perturbations in the fixed background with an interesting observation
regarding the non-diagonal perturbation equations. They reveal an
interesting feature of the present problem. From the form of Eq.~(\ref{nondiag})
one can see that each of the equations contain one time derivative. As
a result both $\delta u_i$ and $\delta F_c$ cannot have time dependent
parts as $e^{i\lambda t}$ for a real $\lambda$. This shows that the
perturbations in the fluid velocity cannot be oscillating in time as 
$e^{i\lambda t}$ and consequently the perturbations must be either
growing or decaying with time or in other words the perturbations are
unstable. This fact may not be a completely unexpected as we have seen
that the background solution predicts a fluid with negative pressure
throughout space and it is well known that perturbations in a
negative pressure fluid are prone to be unstable as pressure forces
cannot withstand the force of fluid compression.  The unstable
perturbations in the present system can be of various kinds, some of
which does not affect the energy density while others can affect
energy density and pressure. Perturbations growing in time may
ultimately destroy the background set up where as perturbations
decaying in time will be unable to destabilize the background fluid
properties.
\section{Conclusion}

In this article we have applied the modified Einstein equations,
arising out of non-minimal coupling of a perfect fluid with gravity,
in the particular case of the Schwarzschild spacetime. Knowing the
metric solution we derive the properties of the fluid in such a
spacetime. Although Schwarzschild spacetime has $R=0$, but the effects
of the NMC, introduced in the fluid-gravity action, remains in the
modified Einstein equation.  The very fact that a non-minimal coupling
term in the gravity-fluid Lagrangian can affect the solutions of the
modified Einstein equation for the Schwarzschild spacetime, for which
$R=0$, is interesting.  In this article we have also presented the
solution of the modified Einstein equation for the flat
spacetime. There also we find an interesting non-trivial solution
which predicts that in presence of NMC there can be a fluid in flat
spacetime, whose energy-density and pressure are constants. The
important point about the solution is that the pressure of such a
fluid must be negative. The flat spacetime solution is important for
the present work as Schwarzschild spacetime is asymptotically flat,
and consequently all the large $r$ behavior of the fluid present in
the Schwarzschild spacetime must merge with the flat spacetime
results.

In this article we have assumed $F_c$ to be like an independent scalar
field, which depends on the fluid parameters, and plays the most
important role for non-minimal coupling of the fluid with gravity. Our
method always works when one chooses those spacetimes for which $R=0$.
At the end we want to stress that although we gave a slightly
different interpretation of $F_c$, it still retains all its basic
properties. For the time-independent and radially symmetric solutions,
one can easily check from the $1-1$ component of the modified Einstein
equation in the Schwarzschild spacetime that if there is no ambient
fluid, that is if $F=0$ (and consequently $n=0$), one gets $F_c$ to be
a constant. From the form of the fluid-gravity action, as given in
Eq.~(\ref{nmcact}), it can be checked that in such a case one gets
back the Einstein equation of GR in vacuum and the effect of NMC gets
completely wiped out as expected.  This implies that non-trivial
effects of NMC arises, through the effect of $F_c$, only when there 
is an ambient fluid in the curved spacetime. In the finite $r$ regions
of the Schwarzschild spacetime one can write $F_c(n)$ by using the
form of $n(r)$ in Eq.~(\ref{nr}). The fact that $F_c$ depends on the
radial coordinate, for the case of flat spacetimes, is directly
related to the fact that in flat spacetime $n$ is a constant. If $F_c$
remains purely a function of $n$ and hence becomes a constant when
$F\ne 0$, then one gets an inconsistent Einstein equation from the
form of the action in Eq.~(\ref{nmcact}). For a relevant form of $F_c$
in flat spacetime it must depend upon some coordinates, and in a
spherically symmetric situation it is natural that it depends upon the
radial coordinate $r$.  More over as we were working with a static
fluid in a time-independent set up, one can trivially show that the
functional form of $n(r)$ satisfies the constraint $\nabla_\mu (n
u^\mu)=0$. So one can indeed interpret $n(r)$ as a number density
specifying the fluid.

In the Schwarzschild spacetime it is seen that one can always
accommodate a perfect fluid whose nature changes radially. In the
region $2\le r < 3$ the fluid is a phantom fluid and it becomes a dark
fluid with $-1\le (p_{\rm tot}/\rho_{\rm tot}) \le -(2/3)$ in the
region $r \ge 3$. One can keep the energy-density of the perfect fluid
to be positive in the physically accessible regions of spacetime. The
interesting thing to note is that the perfect fluid, non-minimally
coupled to gravity in the Schwarzschild spacetime, must have negative
pressure. One cannot have non-minimally coupled positive pressure
fluid in the Schwarzschild spacetime. This is very different from the
case of minimally coupled perfect fluid in the Schwarzschild
spacetime, where one can have all sorts of perfect fluids which have
zero, positive or negative pressure. In this article we have analyzed
the region of spacetime for which $r>2$, outside the event horizon as
in the Schwarzschild spacetime that is the region accessible to an
observer at infinity.

In the case of non-minimally coupled fluid one sees that the fluid has
negative pressure throughout but in regions $r>3$, the quantity
$\rho_{\rm tot} + p_{\rm tot}>0$ and only in the region $r<3$ one has
$\rho_{\rm tot} + p_{\rm tot}<0$ and the pressure diverges in the
negative direction where as the energy-density tends to zero near the
event horizon. The expression of the energy-density and pressure of
the fluid shows that even if $n=0$ the energy-density and pressure
does not turn out to be zero. This is a very interesting property of
the non-minimally coupled fluid in the Schwarzschild spacetime. The
$n$ independent part of the fluid parameters specify the dark energy
like contributions. In the region $r>3$, decrease in $n$ is
accompanied by decrease in $\rho_{\rm tot}$, but once in the region
$2<r<3$, $\rho_{\rm tot}$ decreases as number density increases. In
the region $2<r<3$ the energy-density $\rho_{\rm tot}(n)$ is a
decreasing function of $n$ because in this region $\rho_{\rm tot} +
p_{\rm tot}<0$.  From the expressions of energy-density and pressure
in Eq.~(\ref{rhopt}) one can verify that when $\rho_{\rm tot} + p_{\rm
  tot}<0$ one must have $({\partial{\rho_{\rm tot}}}/{\partial{n}})<0$
as number density $n(r)$ cannot be negative. This interesting behavior
of energy-density can also be obtained in minimally coupled fluids in
GR when the weak energy condition is violated, as can be checked from
the expression of pressure in Eq.~(\ref{rhop}). In our particular case
as $\rho_{\rm tot} + p_{\rm tot}$ is positive for $r>3$ and negative
for $r<3$ attaining zero value at $r=3$ where we have
$n({\partial{\rho_{\rm tot}}}/{\partial{n}})=0$. At $r=3$ it is seen
that $({\partial{\rho_{\rm tot}}}/{\partial{n}})$ is discontinuous but
not singular and so $n$ becomes zero at $r=3$. We have seen that for
the background solution the pressure diverges at the horizon. One must
note that pressure is a scalar function and consequently a divergence
in its value shows a physical effect independent of any coordinate
choice. The divergence of the pressure function at the horizon
therefore is a physical effect and not due to some wrong choice of
coordinates.

The present observations regarding the existence of non-minimally
coupled dark fluid in presence of the Schwarzschild spacetime turns
out to be interesting as one can pinpoint most of the properties of the
fluid from the modified Einstein equation. In this paper we assumed
that the fluid is at rest in the Schwarzschild coordinates. One can
also get back the vacuum solutions in the present case by suitably
choosing the constants appearing in the analysis. The present paper
shows that simplest of the blackholes can be surrounded by an exotic
perfect fluid whose property changes with radial distance. More over
this fluid does not vanish at large distance from the event horizon,
the energy-density and the pressure of the fluid becomes constant in
the large $r$ limit. As because the system is made up of negative
pressure fluid the fluid perturbations on the system turn out to be
unstable. Some perturbations may lead to distabilization of the
background solutions. In the case of minimally coupled fluid, one has
to always assume that the fluid is not backreacting on the spacetime,
an assumption which is at best an approximation because any fluid
which has high enough energy-density or pressure may actually modify
the original spacetime itself. Unlike the minimally coupled case, the fluid 
properties in the case of non-minimal coupling, are obtained using the
proper modified Einstein equation and consequently the solutions are
always valid and one may not treat the fluid as a test probe in the
Schwarzschild spacetime.
\vskip 1cm
\noindent
{\bf Acknowledgement:} The authors thank Sayantani Bhattacharya, in
the physics department of Indian Institute of Technology, Kanpur, for
helpful discussions.


\end{document}